# Role of quasi-Fermi levels in Si- and Mg-related optical absorption in nitride laser diodes (LDs): material context

Konrad Sakowski[1,2,*], Cyprian Sobczak[1], Paweł Strak[1], Izabella Grzegory[1], Robert Czernecki[1], Jacek Piechota[1], Agata Kaminska[1,3,4], Stanislaw Krukowski[1]

[1]*Institute of High Pressure Physics, Polish Academy of Sciences, Sokolowska 29/37, 01-142 Warsaw, Poland*

[2]*Institute of Applied Mathematics and Mechanics, University of Warsaw, 02-097 Warsaw, Poland*

[3]*Faculty of Mathematics and Natural Sciences. School of Exact Sciences, Cardinal Stefan Wyszynski University in Warsaw, Dewajtis 5, 01-815 Warsaw, Poland*

[4]*Institute of Physics, Polish Academy of Sciences, Aleja Lotnikow 32/46, PL-02668 Warsaw, Poland*

[*] *Corresponding author, e-mail: konrad@unipress.waw.pl*

Optical absorption and reabsorption of light emitted from active regions in nitride laser diodes (LDs) have been shown to reduce the light extraction efficiency of these devices. It was proven that the presence of Si and Mg may considerably increase the optical absorption. This effect is much stronger in the high-energy (short-wavelength) range of the spectrum. The absorption increase is directly related to the ionization of the Si donor and Mg acceptor levels, which are controlled by the electron and hole quasi-Fermi levels. It is shown that the absorption may be increased because of the higher ionization of Mg caused by the compensation in the p-type region and the high ionization of Si in the n-type region. It was explained theoretically why optical efficiency is increased by removal of doping in waveguides. It was also shown that good material quality leads to a low absorption level, especially in the Mg-doped p-type part of the device.

**Keywords:** gallium nitride, laser diodes, magnesium, optical absorption, doping

# I. Introduction

An increase in the optical efficiency of nitride devices has led to significant progress in the development of nitride optoelectronics [1, 2]. Nevertheless, there is still considerable room for improvement, both in terms of the electric and optical properties of devices, such as light-emitting diodes (LEDs) [1, 2] and laser diodes (LDs) [3]. This may also lead to an extension of the emission spectrum [4, 5]. There are many ways to further increase the optical output of the device. It is particularly rewarding that progress in parallel developments may be additive, thus enhancing their contribution to overall improvement.

In recent years, significant conceptual progress has been achieved in the electric properties of nitride devices [6, 7]. The most formidable barriers in electrical design were systematically studied, creating a path for their removal. These barriers are related to the p-type region: the low electric conductivity and high resistance of the metal-p-type contact [2, 3, 4, 5, 6, 7, 8]. Low p-type conductivity is related to the high activation energy of the only effective acceptor, magnesium, such that about one percent of Mg only is occupied at room temperature [9]. As a result, the hole density is low, which, in conjunction with the low hole mobility, creates serious difficulties. This remedy was proposed by the invention of polarization doping, which postulated the emergence of a polarization-related electric charge that may induce a hole-rich region [10, 11]. This was not in the stage of working design, but recently, the concept of polarization and mobile charge was formulated, which led to mobile hole density without the application of magnesium [12]. The second hurdle is related to p-type contact, which is caused by the high ionization energy of nitrides, that is, the low energy of the valence band states. Recently, a full analysis of the p-type contact was performed, followed by a multilayer design incorporating a number of deep acceptors to overcome the Schottky barrier [13]. This design is yet to be implemented; nevertheless, it is hoped that an electrically effective nitride device will be created in the near future.

Other features are related to the optical properties of the devices, which are also improved. This was achieved for both light emitting diodes and laser diodes. Nevertheless, there is still room for improvement, so some features still need further work. One of these aspects is light extraction from the laser diode, which is discussed in the present work. An important feature is the reabsorption of the emitted light, which may be partially related to the presence of ionized defects intentionally introduced in the device: Si donors and Mg acceptors. These problems are analyzed below.

The remainder of this paper is organized as follows. First, the device design is discussed, including the general features and the specific design of the device investigated. Subsequently, the model is presented, which is used in the calculation of the basic characteristics. The main results are then presented and discussed briefly. Finally, the results are briefly summarized.

## II. The device design

Over the years of development, the design of the laser diode was modified considerably. This has led to a large number of different laser diode types, which require extended studies. The basic features of the device remained the same as those in the original design of Nakamura in Nichia Chemical labs [3]. The diode consists of a variable number of layers, as cladding layers can be realized as superlattices consisting of a large number of chemically different layers. The simplest design is a p-n diode with multiquantum wells (MQWs) working as the active part of the device, which is composed of several narrow layers of GaInN and GaN. The active layers were topped with a thin GaN layer. Below is the waveguide, n-type cladding, and sub-contact-n layer. Above, there is a p-type doped electron blocking layer, the second waveguide, the cladding, and the subcontact-p at the top. The details vary as both waveguides may be doped. The EBL and top cladding were p-type-doped. The subcontact layers are both doped and most likely processed by additional annealing of the p-type contact.

The details of the design vary depending on the growth method and the substrate. For example, EBL may be positioned close to MQWs, as in the early designs, or shifted to the cladding, as has been made more recently. Other frequent changes include doping and thickness of the wells and barriers. For demonstration purposes, we present three examples of laser diodes. In cases ($Case$ 1, $Case$ 2, $Case$ 3) differ in the material properties differed only (doping), and the geometric design was the same.

$Case$ 1 resembles a relatively early design, which is a modification of Nakamura et al.'s original design [3]. The well/barrier regions are always doping-free. The EBL was shifted from the neighborhood of the wells to the p-type waveguide-subcontact-p interface. Several factors have contributed to this shift:

(i) The removal from the proximity of the wells so that Mg could possibly be less detrimental as it was positioned at some distance.
(ii) The danger of Mg diffusion into the wells was experimentally shown to kill the emission from the device.
(iii) The stability of the In wells was improved as high-temperature EBL growth occurred after a thick GaN layer (waveguide) was grown.

The n- and p-type doping was also used in the waveguide layers, that is, over subcontacts, claddings, and waveguides. The p-type regions had compensation of 10%. Thus, the doping-free layer was relatively small and limited to the well/barrier regions. $Case$ 2 differed by the removal of doping in both the waveguides. Thus, the doped regions include both the subcontacts and cladding. The p-type regions had compensation of 10%. Finally, $Case$ 3 differs from $Case$ 2 in the absence of compensation in the p-type.

These three cases are presented in Table 1. They differ only by doping. When doping is different between the cases, the upper, middle, and lowest entries refer to $Case$ 1, $Case$ 2 and $Case$ 3, respectively. When these cases were identical, a single value was obtained.

Table 1. Blue laser structures, used in simulations. Most of the parameters is the same for all three considered cases. If this is not the case, the upper, middle, and lowest entries in the table refer to $Case$ 1, $Case$ 2 and $Case$ 3, respectively.

| Layer | Thickness [$nm$] | Chemical composition | Si donor density [$cm^{-3}$] | Mg acceptor density [$cm^{-3}$] |
|---|---|---|---|---|
| Subcontact - n | 30 | $GaN$ | $3 \times 10^{18}$ | 0 |
| Cladding | 300 | $GaN$ | $2 \times 10^{18}$ | 0 |
| Waveguide | 150 | $Ga_{0.9}In_{0.1}N$ | $2 \times 10^{18}$ <br> 0 <br> 0 | 0 |
| Well | 3 | $Ga_{0.8}In_{0.2}N$ | 0 | 0 |
| Barrier | 8 | $Ga_{0.9}In_{0.1}N$ | 0 | 0 |
| Well | 3 | $Ga_{0.8}In_{0.2}N$ | 0 | 0 |
| Barrier | 8 | $Ga_{0.9}In_{0.1}N$ | 0 | 0 |
| Waveguide | 150 | $Ga_{0.9}In_{0.1}N$ | $2 \times 10^{18}$ <br> 0 <br> 0 | $2 \times 10^{19}$ <br> 0 <br> 0 |
| Electron blocking layer | 20 | $Al_{0.9}Ga_{0.1}N$ | $2 \times 10^{18}$ <br> $2 \times 10^{18}$ <br> 0 | $2 \times 10^{19}$ |
| Cladding | 300 | $GaN$ | $1 \times 10^{18}$ <br> $1 \times 10^{18}$ <br> 0 | $1 \times 10^{19}$ |
| Subcontact - p | 30 | GaN | $3 \times 10^{18}$ $3 \times 10^{18}$ 0 | $3 \times 10^{19}$ |

Similar structures were used in the fabrication for the optimization of the device design, grown for blue-violet laser diodes, and used in many laboratories. It is assumed that at both ends, the mirrors are made of distributed Bragg reflectors (DBRs). It is assumed that the front mirror has reflectivity $R_1 = 0.2$ and the rear mirror has reflectivity $R_2 = 0.8$. Thus, the dominant portion of the light escapes from the device in the first one, as the transmission coefficient is $T_1 = 1 - R_1 = 0.8$. Nevertheless, some light travels back and forth many times before final escape, thus absorption in the bulk is important factor affecting the critical parameter, i.e. light extraction efficiency $\eta_{ex}$.

### III. The model and calculations

The efficiency of the optoelectronic device was measured by the wall plug efficiency, that is, the ratio of the energy of the emitted light to the electric energy alimenting the device. Despite the simplicity of this parameter, determining its value in some cases may be cumbersome. In fact, this is more complex in the case of LEDs, as light is emitted in every direction, which is difficult to measure. Laser diodes are relatively simple in that matter.

The wall plug efficiency is degraded by a large number of factors, each of which swallows some portion of the total energy input. Fortunately, they are additive; therefore, they can be considered separately. The latter is light extraction, which is related to radiative recombination and subsequent reabsorption and may capture a significant part of the emitted light. The influence of different factors is described by the rate equation for photons of mode $m$ of the edge-emitting laser diode [14]:

$$\frac{dS_m}{dt} = \left(\frac{c}{n_{eff}} G_m - \frac{1}{\tau_{ph}}\right) S_m + C_m R_{sp} \qquad (1)$$

where $S_m$ is the integral photon density obtained by averaging the photon density $s_m(x,y)$ in the $(x,y)$ planes perpendicular to the mode $m$ propagation, $G_m$ and $C_m$ are the gain and spontaneous emission factors, respectively. The photon lifetime $\tau_{ph}$ encompasses the photon "loss" due to absorption (absorption coefficient $\alpha$) and reflectivity at the front and rear facets ($R_1$, $R_2$):

$$\frac{1}{\tau_{ph}} = \frac{c}{n_{eff}}\left[\alpha_m - \frac{1}{2L}\ln(R_1 R_2)\right] \quad (2)$$

where $L$ is the cavity length, and $n_{eff}$ is the effective refractive index ($n_{eff} = 2.5$ for GaN [15][14]). From these data, the light extraction efficiency $\eta_{ex}$ may be obtained for application to a laser diode in analogy to LEDs [16, 17]:

$$\eta_{ex} = exp\left(-\frac{\alpha L}{2}\right)(1 - R_1)(1 + Q + Q^2 + \cdots) \quad (3a)$$

where the first factor stems from the fact that, on average, the starting emission position is uniformly distributed over the cavity length $L$, the factor $T_1 = 1 - R_1$ is the transmission across the front mirror, and the factor $Q$ is the light loss in one full cycle over the cavity, which is

$$Q = R_1 R_2 exp(-2\alpha L) \quad (3b)$$

That gives the final result:

$$\eta_{ex} = exp\left(-\frac{\alpha L}{2}\right)\frac{(1-R_1)}{1 - R_1 R_2 exp(-2\alpha L)} \quad (3c)$$

In the following, we assume that the optical wave is uniformly distributed in the waveguide; thus, we can average over the entire area by employing the integral photon density $S_m$ and effective refractive index $n_{eff}$ in Eq. 1.

Naturally, the emission energy from In-containing wells is lower than the GaN bandgap, thus preventing direct absorption. Nevertheless, in the presence of Mg acceptors or Si donors, a direct transition from the acceptor level to higher-energy conduction band states is possible [18, 19]. The absorption coefficient $\alpha$ is equal to [19]:

$$\alpha(N_D, N_A, x, y) = a_1(N_D, N_A) \exp\left[\frac{h\nu - E_g(x,y)}{b_1(N_D, N_A)}\right] + a_2(N_D, N_A) \exp\left[\frac{h\nu - E_g(x,y)}{b_2(N_D, N_A)}\right] \quad (4)$$

were found to be in reasonable agreement with the experimental data. The densities of the Si donor and Mg acceptors were $N_D$ and $N_A$, respectively (given in $10^{18}\ cm^{-3}$). The scaling constants were adjusted to those of nitrides [19].

$$a_1[cm^{-3}] = 19000 + 4000\ N_D + 200\ N_A \quad (5a)$$
$$b_1[cm^{-3}] = 0.019 + 0.001\ N_D + 0.0001\ N_A \quad (5b)$$
$$a_2[cm^{-3}] = 330 + 200\ N_D + 30\ N_A \quad (5c)$$

$$b_2[cm^{-3}] = 0.07 + 0.016\, N_D + 0.0008\, N_A \quad (5d)$$

These parameters are expressed as: $a_i[cm^{-1}]$ and $b_i[eV]$. Thus, the transition probability is proportional to the occupation probability of the Mg level, which is given by the Fermi-Dirac distribution function for holes. At equilibrium, this is parameterized by the Fermi level common for electrons and holes. Optical absorption may be related to various transitions, such as the direct acceptor to conduction band transition or the Auger effect [15]. These effects were not specified in Ref. [19], thus it is assumed that the experimental data include all the contributions. Hence, it is assumed that the derived analytical expressions in Eqs 4 and 5 include all processes. In contrast, the transition rate is proportional to the occupation of the Mg acceptor level.

$$\alpha(E_F, x, y) = M f_{FD}(E_{Mg}) = M\, N_A / \{1 + g_A exp[(E_A - E_F)/kT]\} \quad (6)$$

where $g_A = 4$. Similarly, the electron after the transition may land at the empty Si donor level:

$$\alpha(E_F, x, y) = M N_D^+ = M N_D [1 - f_{FD}(E_{Si})] = M\, N_D / \{1 + exp[(E_F - E_D)/kT]\} \quad (7)$$

where M denotes the oscillator strength. Therefore, the absorption coefficient must be recalculated using the above relations and the positions of the quasi-Fermi levels of the electrons and holes:

$$\alpha(N_D, N_A) = \alpha(N_D = 0, N_A = 0) + \Delta\alpha(N_D) + \Delta\alpha(N_A) \quad (8)$$

where $\alpha(N_D = 0, N_A = 0)$ are the values obtained from Eqs 4 and 5, and the donor contribution is related to the density of ionized Si donors (empty):

$$\Delta\alpha(N_D) = [\alpha(N_D, N_A = 0) - \alpha(N_D = 0, N_A = 0)] \frac{N_D^+(E_F^n)}{N_D^+} \quad (9)$$

where $N_D^+(E_F^n)$ and $N_D^+$ are the numbers of ionized donors in the device for the quasi-Fermi level $E_F^n$ and in the Si-doped GaN bulk, respectively, both doped to the donor density $N_D$. A similar proportionality may be used for the adsorption by ionized (occupied) Mg acceptors:

$$\Delta\alpha(N_D) = [\alpha(N_D = 0, N_A) - \alpha(N_D = 0, N_A = 0)] \frac{N_A^-(E_F^p)}{N_A^-} \quad (10)$$

where $N_A^-(E_F^p)$ and $N_A^-$ are the numbers of ionized donors in the device for the quasi-Fermi level $E_F^p$ and in the Mg-doped GaN bulk, respectively, both doped to the acceptor density $N_A$. The above equations can be used to determine the optical absorption in nitride-based LD structures, obtained from the solution of the van Roosbroeck drift-diffusion equations [20]. It was assumed that the system was uniform along the diode and that the transversal dependence was not critical. Thus, the problem was reduced to a one-dimensional set of differential equations:

$$-\nabla \cdot \left(\varepsilon_o \varepsilon_{33}(z) \nabla V(z)\right) = p(z) - n(z) - N_A^-(z) + N_D^+(z) - \nabla P_0 \quad (11a)$$

$$-\nabla \cdot \left(\mu_n n(z) \nabla E_F^n(z)\right) = -eR \quad (11b)$$

$$-\nabla \cdot \left(\mu_p p(z) \nabla E_F^p(z)\right) = eR \quad (11c)$$

where the electron and hole densities are calculated in the nondegenerate band approximation:

$$n(z) = N_C exp\left[\left(\left(E_F^n(z) - E_C(z) + eV(z)\right)/kT\right)\right] \quad (12a)$$

$$p(z) = N_V exp\left[\left(\left(E_F^n(z) - E_V(z) - eV(z)\right)/kT\right)\right] \quad (12a)$$

$N_C$ and $N_V$ and $E_C$ and $E_V$ are the density of states energies of the conduction and valence bands.

## IV. The results

At equilibrium, the Fermi level was uniform across the device. The built-in electric field defect level may be shifted, particularly close to the interfaces. Rather, regions of high fields may be affected. As shown in Fig. 1 2 and 3, the results were different, indicating the crucial impact of doping. This is illustrated by the solution of the drift-diffusion set of equations for a typical nitride laser diode, which is active in the blue range. The structures presented in Table 1 were calculated using the drift-diffusion model based on a system of van Roosbroeck equations [20, 21]. The system was solved using the composite discontinuous Galerkin method [22, 23, 24]. The results are shown in the band profile of the device at zero voltage and operating at 4 V. In Fig 1, the *Case* 1 is presented.

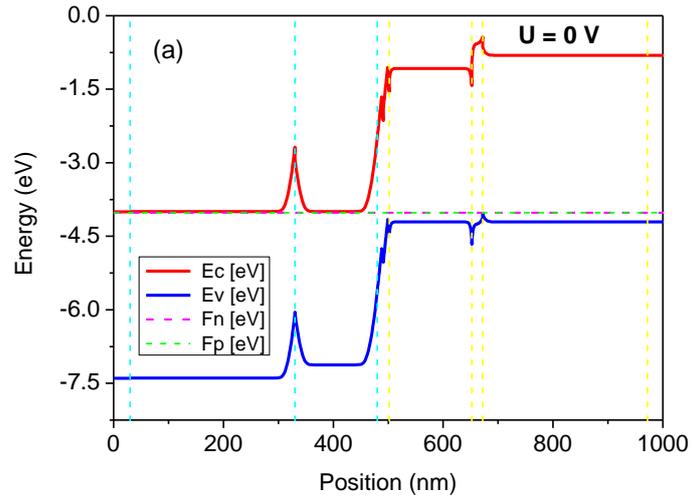

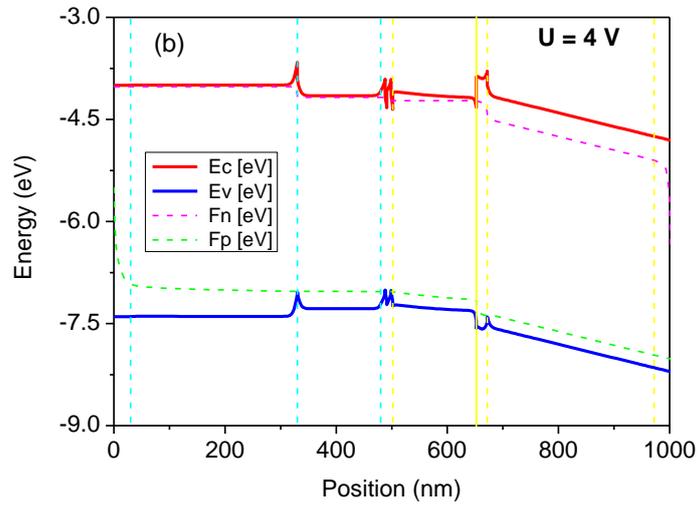

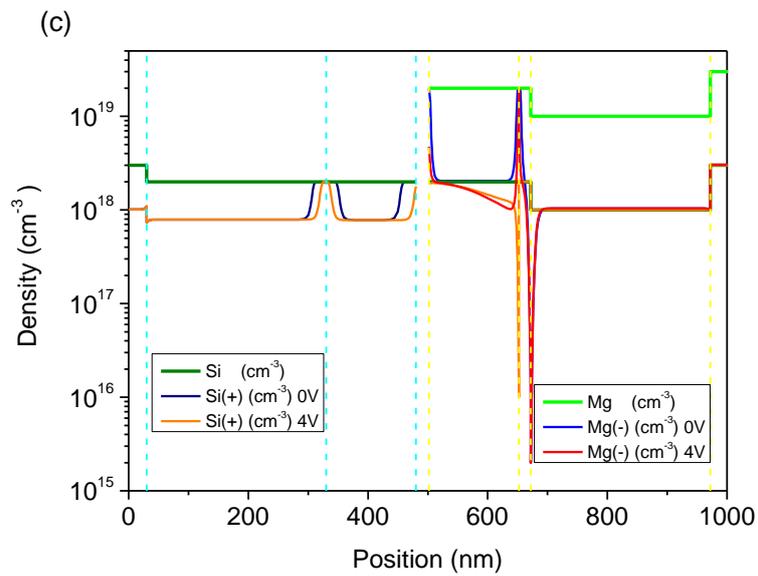

*Fig. 1. Electric properties of nitride based laser diode* Case *1 design, obtained from solution of the set of drift-diffusion equations: band profiles: (a) U = 0 V, (b) U = 4 V, (c) total and ionized defect density for both voltages. Red and blue solid lines represent conduction band minimum (Ec- CBM) and valence band maximum (Ev -VBM), respectively. Dashed magenta and green lines denote quasi-Fermi levels for electrons (Fn) and holes (Fp), respectively. The vertical dashed lines mark layers with n-type (cyan) and p – type (yellow) doping. The layers without doping are not marked. The total and ionized defects are marked by different colors.*

From these results, it follows that doping plays an important role by screening the field-limiting regions of high ionized defect density close to the heterointerfaces. This is especially important in EBL. The p-n junction field, which is present at zero voltage, is located close to the wells. An additional change in the conduction band is visible in the EBL, owing to the bandgap difference. The valence band is controlled by the Fermi level and acceptor ionization energy, which were first used by Nakamura in his new design of nitride devices [3].

The region of high concentration of ionized Mg acceptors is limited to part of the EBL, whereas a high concentration of ionized Si donors is at the cladding-waveguide interface and the p-n junction. In fact, the region of fully ionized Si donors (Schottky regime), that is, $N_D^+ = 2 \times 10^{18}\ cm^{-3}$ is 30 nm wide for $U = 0\ V$ and significantly reduced for $U = 4\ V$. The region of reduced concentration of Si donors to $N_D^+ = 1.3 \div 1.9 \times 10^{18}\ cm^{-3}$ is only 10 nm wide. Therefore, the influence of ionized donors is reduced when laser diode is operating. The variation in the concentration of ionized Mg acceptors was much larger. For the cladding, the density of the ionized Mg acceptor is $N_A^- = 2.0 \times 10^{18}\ cm^{-3}$, which is essentially equal to the density of compensating donors. The much larger density is at the EBL edge, which is $N_A^- = 2.0 \times 10^{19}\ cm^{-3}$, that is, almost full ionization of acceptors of the density $N_A = 2 \times 10^{19}\ cm^{-3}$. This is narrow, approximately 5 nm wide, for $U = 0\ V$.

In the following, the design is modified by the removal of doping from the lower waveguide. This leads to some drastic consequences, which are most visible at zero voltage, as shown in Fig. 2.

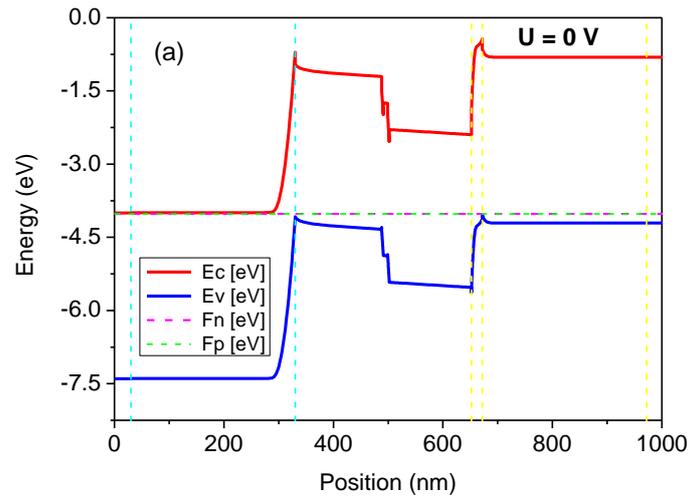

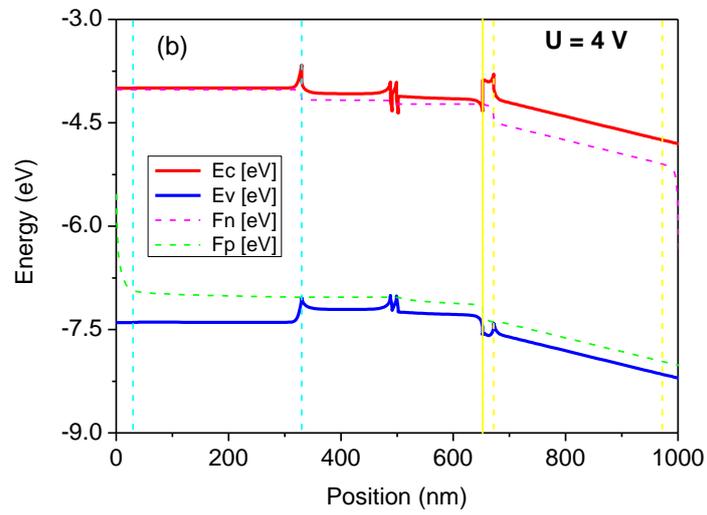

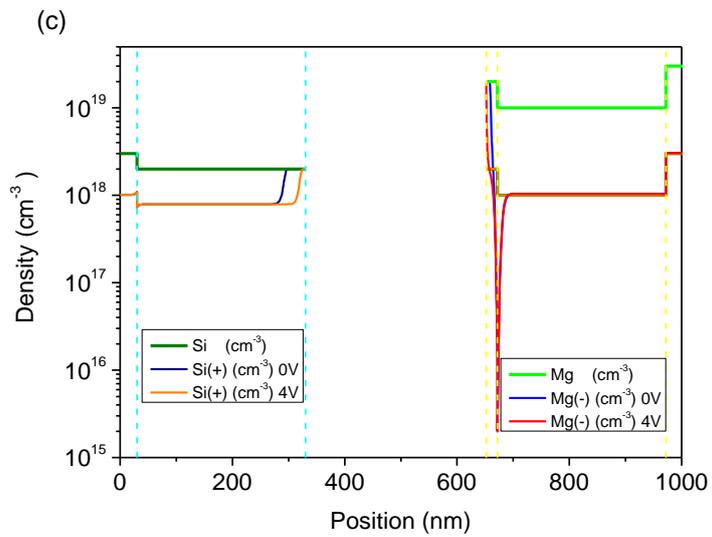

*Fig. 2. Electric properties of nitride based laser diode Case 2 design, obtained from solution of the set of drift-diffusion equations: band profiles: (a) U = 0 V, (b) U = 4 V, (c) total and ionized defect density for both voltages. The symbols used are denoted as in Fig. 1.*

The junction field was split into two parts: n-type cladding and EBL. This is due to the fact that screening is mostly due to ionized donors, so they are available separately. Thus, as before, the ionized screening layer was split with no screening close to the wells. In the case of zero voltage, an additional opposite-directed field emerges in the active region, which considerably hampers emission. The application of the voltage $U = 4\,V$ leads to the disappearance of the field and reduction of the ionized layer, both in terms of density and thickness.

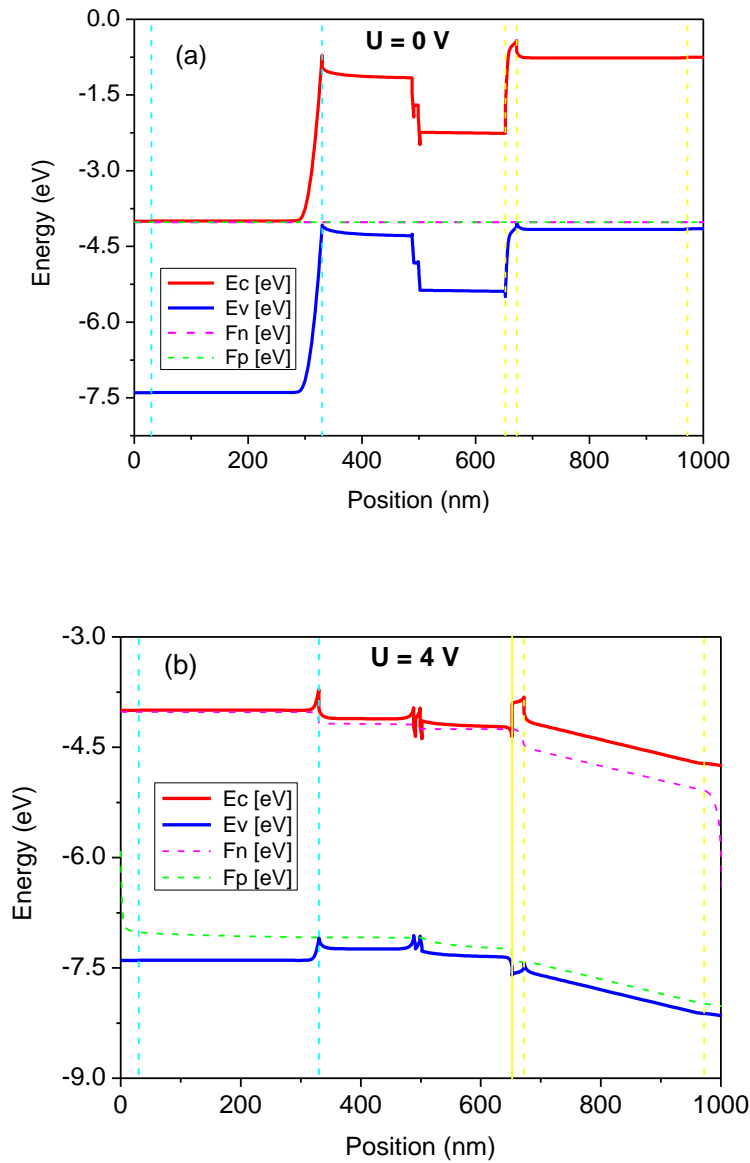

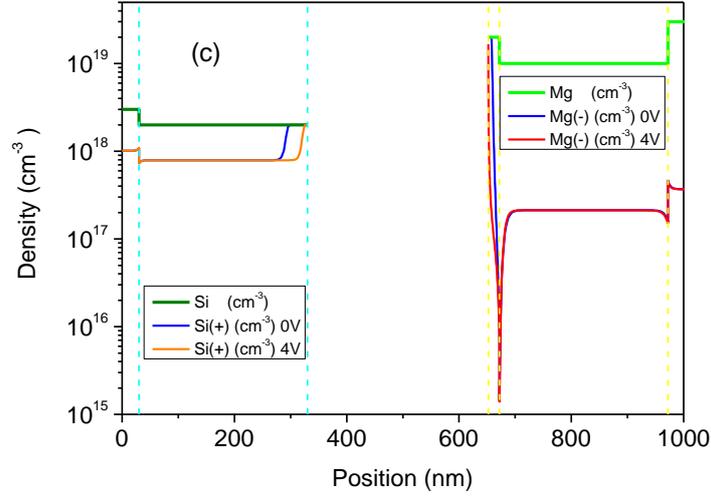

*Fig. 3. Electric properties of nitride based laser diode* Case *3 design, obtained from solution of the set of drift-diffusion equations: band profiles: (a) U = 0 V, (b) U = 4 V, (c) total and ionized defect density for both voltages. The symbols used are denoted as in Fig. 1.*

Finally, the influence of compensation in the p-type was investigated by its removal in the p-type region of the device. The band profile shows that the p-n junction is separated into two parts: at the cladding – waveguide interface and the EBL, as the waveguide was not doped. These jumps vanished at a voltage $U = 4\,V$. These regions are strongly affected by the electric field, which can induce a higher density of charged defects. In addition, during device operation, a non-equilibrium between electrons and holes is created.

In this design, the n- and p-type regions are separated and limited to the part of the device. In most of the volume, the fraction of ionized defects is small, equal to the bulk values i.e. about 10% and 1% in case of Si and Mg, respectively. The region of high concentration of ionized Mg acceptors is limited to part of the EBL, whereas a high concentration of ionized Si donors is only at the cladding-waveguide interface. In fact, the region of fully ionized Si donors (Schottky regime), that is, $N_D^+ = 2 \times 10^{18}\,cm^{-3}$ is 30 nm wide for $U = 0\,V$ and essentially vanishes for $U = 4\,V$. The region of reduced concentration of Si donors to $N_D^+ = 1.3 \div 1.9 \times 10^{18}\,cm^{-3}$ is only 10 nm wide. Therefore, the influence of ionized donors is reduced when operating laser diode. The variation in the concentration of ionized Mg acceptors was much larger. For the cladding, the density of ionized Mg acceptor is $N_A^- = 2.1 \times 10^{17}\,cm^{-3}$, jus approximately 1% of the total density of Mg acceptors $N_A = 2 \times 10^{19}\,cm^{-3}$. The much larger density is at the EBL edge, which is $N_A^- = 2.0 \times 10^{19}\,cm^{-3}$, that is, almost full ionization of acceptors of the density $N_A = 2 \times 10^{19}\ cm^{-3}$. This is narrow (approximately 5 nm wide) for $U = 0\ V$. This is even smaller for LD at an operation voltage $U = 4\ V$. In general, the application of a voltage reduces the number of ionized donors and acceptors. The optical absorption can be obtained using the calculated ionized defect density via Eqs 2 – 12. The absorption coefficients for these three devices are presented in Fig. 4.

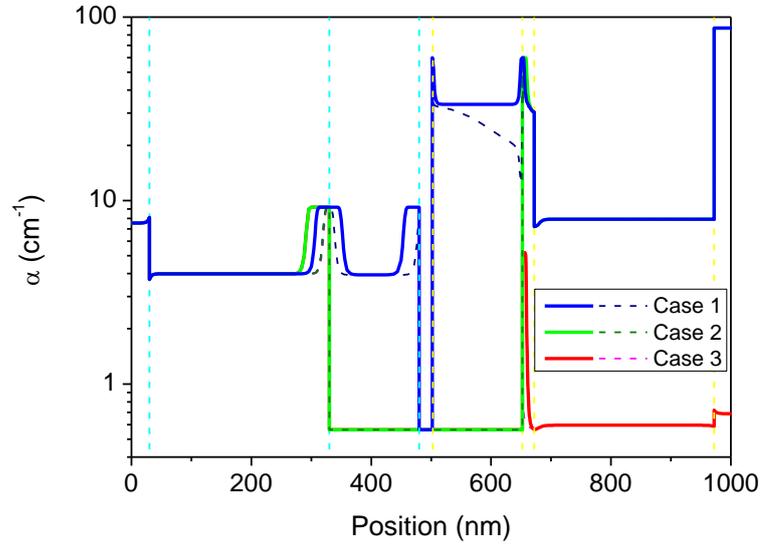

*Fig. 4. Optical absorption coefficient across the nitride laser diode for three designs operating at the wavelength $\lambda = 410\ nm$. The solid and dashed lines are obtained for the two voltages applied to device: U = 0 V and U = 4V. The vertical dashed lines mark borders of the doped regions as above.*

The obtained absorption coefficient $\alpha$ values depend significantly on the concentration of the donors and acceptors. Additionally, increased absorption values were obtained at high concentrations of ionized defects. As this may be controlled by compensation, the important factor is to keep the compensation level value below 1% for p-type. In the n-type part, the primary factor is the Si donor concentration. Much less important is compensation in n-type doped regions because the donor ionization is high.

In the subcontact-n layer, the absorption coefficient is $\alpha = 7.55\ cm^{-1}$ which is reduced to $\alpha = 3.99\ cm^{-1}$ in n-type cladding. It should be noted that both regions contributed considerably to the overall optical absorption of the device. The undoped regions have a very small absorption, which is $\alpha = 0.56\ cm^{-1}$. In fact, the mere presence of Mg in p-type cladding increases the absorption only slightly to $\alpha = 0.59\ cm^{-1}$. This is due to the very small fraction of ionized Mg acceptors. An increase in Mg doping in the subcontact-p layer does not increase significantly, and the adsorption there is only $\alpha = 0.69\ cm^{-1}$. An exception to this rule is the absorption in EBL, which achieves $\alpha = 5.20\ cm^{-1}$. The EBL role was reduced because of the relatively narrow layer containing ionized Mg acceptors. The Different values are obtained for the case of p-type 10% compensation: for EBL absorption value increases to $\alpha = 50.0\ cm^{-1}$, in subcontact-p, it is at very high value: $\alpha = 87.1\ cm^{-1}$. A much smaller value was observed in the p-type cladding: $\alpha = 7.91\ cm^{-1}$. Despite the reduced value, this is considerably higher than that in the pure Mg doping, which is one order of magnitude smaller, that is, $\alpha = 0.59\ cm^{-1}$. In the p-type waveguide, it attains $\alpha = 39.71\ cm^{-1}$ a considerable effect. Therefore, p-type compensation related absorption is primary detrimental effect, indicating the necessity of maintaining compensation in p-type below 1%, i.e. equilibrium ionization value.

Optical absorption can be used to determine light extraction efficiency coefficient $\eta_{ex}$. In this calculation, it was assumed that the optical wave intensity was uniform across the entire laser diode, which is a good approximation for a typical device. In the analysis, it was assumed that the photons may leave the device at the selected point; therefore, the local value is determined $\eta_{ex}(z)$. The values of the coefficients obtained from Eqs 3 are plotted in Fig 5.

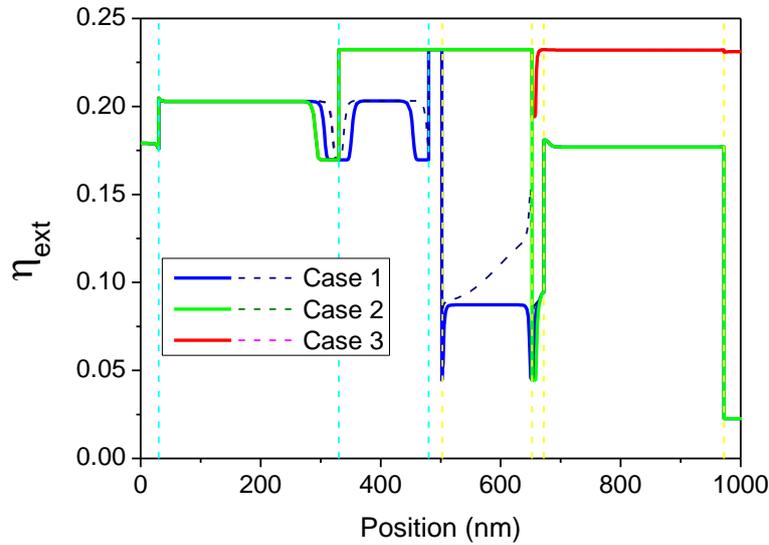

*Fig. 5. Light extraction efficiency across the nitride laser diode for three designs operating at $\lambda = 410\ nm$. The solid and dashed lines present the results obtained for the two voltages applied to device: U = 0 V and U = 4V. The vertical dashed lines mark borders of the doped regions as above.*

Generally, the value of the coefficient $\eta_{ex}$ for $\lambda = 410\ nm$ is within the interval $0.169 < \eta_{ex} < 0.232$, thus such a variation seems to be relatively small. This has to be considered as the relative change that translates into loss of 30 % of the output. It should be noted that this value depends on the doping and compensation. The exception is the n-type part, in which these values are $\eta_{ex} = 0.179$ and $\eta_{ex} = 0.203$ for the subcontact-n and cladding layers, respectively. In EBL, the efficiency was reduced to $\eta_{ex} = 0.194$ for Mg doping only. The drastic reduction is related to compensation; in the p-type cladding, this is reduced to $\eta_{ex} = 0.177$. Even more catastrophic reduction in in the p-type waveguide and subcontact, $\eta_{ex} = 0.092$ and $\eta_{ex} = 0.02$, that is, virtual blockade of the light in these regions.

These values can be summarized by the integral values over the entire device, $\eta_{ex-tot}$. In the absence of doping, $\eta_{ex-tot} = 0.232$. The introduction of uniform doping of Si and Mg at the following levels $N_D = 10^{18}\ cm^{-3}$ and $N_A = 10^{19}\ cm^{-3}$ reduced these values to $\eta_{ex-tot} = 0.210$ and $\eta_{ex-tot} = 0.218$, respectively. This is significant but not destructive to the device. This is in agreement with the device average for $Case$ 1 which was $\eta_{ex-tot} = 0.220$ and $\eta_{ex-tot} = 0.221$ for $U = 0\ V$ and $U = 4\ V$, respectively. The introduction of compensation in $Case$ 2 leads to a reduction to $\eta_{ex-tot} = 0.194$ and $\eta_{ex-tot} = 0.196$ for $U = 0\ V$ and $U = 4\ V$, respectively. An extension of the doping and compensation into the waveguides further reduced these values to $\eta_{ex-tot} = 0.167$ and $\eta_{ex-tot} = 0.172$ for $U = 0\ V$ and $U = 4\ V$, respectively. These values indicate that both the doping level and the compensation may reduce the efficiency by 30%.

It should be noted that these values strongly depend on the emission wavelength. This is related to the stronger absorption at higher energies, that is, shorter wavelengths. For example, the efficiency of the uniform device calculated at $N_D = 10^{18}\ cm^{-3}$ and $N_A = 10^{19}\ cm^{-3}$ at $\lambda = 410\ nm$ is $\eta_{ex} = 0.181$, and for $\lambda = 430\ nm$ it is increased to $\eta_{ex} = 0.223$. In contrast, for a shorter wavelength $\lambda = 390\ nm$ it falls dramatically to $\eta_{ex} = 0.078$. Thus, optical extraction is important for short wavelengths but not for longer wavelengths.

## V.     Summary

The light extraction efficiency of a typical nitride-based laser diode device was analyzed using drift-diffusion equations and experimentally derived optical absorption rates. In the analysis, the fraction of ionized donors and acceptors was used to estimate the overall light-extraction efficiency of the device by determining the contribution of the basic segments of the device. The results obtained in this study can be assessed best by representing (i) the state of the art before publication, (ii) resume the results in the paper, and (iii) the state of the art after publication.

Accordingly, the state of the are before publication was:

(i)   Mg related absorption may cause significant detrimental effect to the efficiency of nitride based LDs.
(ii)  Si related absorption is less important, nevertheless it may reduce the output.
(iii) The absorption detrimental effect is essentially the same for all emission wavelengths.

The resume of the results obtained within this publication:

(i)   The defect related absorption may reduce extraction efficiency up to 30%.
(ii)  The optical absorption is dependent on the defect ionization degree.
(iii) The Mg-related absorption is relatively small owing to low ionization degree, the compensation even at 10% drastically increases absorption in p-type part.
(iv)  Si is ionized to high degree, therefore optical absorption is significant.
(v)   The removal of the dopant from the waveguide increases efficiency of the device considerably.
(vi)  The optimal design would be the complete removal of the defects, introduced both intentionally and nonintentionally.

The state of the art after the publication:

(i)   The contribution of Mg to optical absorption and light extraction efficiency deterioration is relatively small, but the compensation increases to dangerous levels.
(ii)  The contribution of Si to optical absorption and light extraction efficiency deterioration is also important, independent of the compensation.
(iii) The excessive n-type and p-type doping should be avoided.
(iv)  The crucial role is nitride material quality, which should, in principle, reduce compensation to extremely low levels.

**Acknowledgments**

This research was partially supported by Polish National Centre for Research and Development grant LIDER/23/0129/L-10/18/NCBR/2019. This research was carried out with the support of the Interdisciplinary Centre for Mathematical and Computational Modelling at the University of Warsaw (ICM UW) under grant no. GB84-23. We gratefully acknowledge Polish high-performance computing infrastructure PLGrid (HPC Center: ACK Cyfronet AGH) for providing computer facilities and support within computational grant no. PLG/2024/017466.

**Author Declarations**

The authors have no conflicts to disclose.

**Data Availability**

The data that support the findings of this study are available from the corresponding author upon reasonable request.